\documentclass[twocolumn,showpacs,amsmath,amssymb,preprintnumbers,aip,jcp]{revtex4}
\usepackage{graphicx}
\usepackage{dcolumn}
\usepackage{color}
\usepackage{bm}
\usepackage{gensymb}
\begin{document}

\title{Study of morphology effects on magnetic interactions and band gap variations for 3$d$ late transition metal bi-doped ZnO nano structures by hybrid DFT calculations}

\author{Soumendu Datta,$^{1,}$\footnote{Electronic mail: soumendu@bose.res.in} Gopi Chandra Kaphle,$^2$ Sayan Baral$^1$ and Abhijit Mookerjee$^{1,}$\footnote{Visiting Faculty, Department of Physics, Lady Brabourne College, Kolkata}$^{,}$\footnote{Visiting
Distinguished Professor, Department of Physics, Presidency University, Kolkata}}
\affiliation{$^1$Department of Condensed Matter Physics and Material Sciences, S.N. Bose National Centre for Basic Sciences, JD Block, Sector-III, Salt Lake
City, Kolkata 700 098, India \\
$^2$ Central Department of Physics, Tribhuvan University, Kathmandu, Nepal}
\date{\today}

\begin{abstract}
Using density functional theory (DFT) based electronic structure calculations, the effects of morphology of semiconducting nano structures on the magnetic interaction between two magnetic dopant atoms as well as a possibility of tuning band gaps have been studied in case of the bi-doped (ZnO)$_{24}$ nano structures with the impurity dopant atoms of the 3$d$ late transition metals (TM) - Mn, Fe, Co, Ni and Cu. To explore the morphology effect, three different structures of the host (ZnO)$_{24}$ nano-system having different degrees of spatial confinement, have been considered : a two dimensional (2D) nanosheet, an one dimensional (1D) nanotube and a finite cage-shaped nanocluster. The present study employs hybrid density functional theory to accurately describe the electronic structure of all the systems. It is shown here that the magnetic coupling between the two dopant atoms, remains mostly anti-ferromagnetic in course of changing the morphology from the sheet geometry to the cage-shaped geometry of the host systems, except for the case of energetically most stable bi-Mn doping, which shows a transition from ferromagnetic to anti-ferromagnetic coupling with decreasing aspect ratio of the host system. The effect of the shape change, however, has a significant effect on the overall band gap variations of both the pristine as well as all the bi-doped systems, irrespective of the nature of the dopant atoms and provides a means for easy tunability of their optoelectronic properties.
\end{abstract}

\maketitle

\section{\label{sec:intro}Introduction}
The II-VI compound semiconductor ZnO with a direct wide band gap of  3.37 eV and a large excitonic binding energy of 60 meV for the bulk hexagonal wurtzite crystal structure at the ambient temperature and pressure condition,\cite{nmat_orientation} has continued to attract great attention till today, both experimentally and theoretically. The reason for this huge interest in ZnO materials is mainly due to its versatile properties in piezoelectric, optical, magneto-electronics, highly efficient blue light emitting diodes (LEDs) and microwave devices.\cite{zno2} TM-atom doped ZnO was also predicted as one of the most promising materials for room temperature (RT) dilute magnetic semiconductor (DMS)\cite{dms1,dms2,nano_dms} to provide more new functionalities such as spin-based information storage, data processing, spin-polarized laser and so on. Therefore, the TM-atoms doped ZnO as well as other II-VI and III-V semiconductors have also been intensively explored for the spintronics applications.\cite{spintron1,spintron2} Another interesting fact is that ZnO is transparent to visible light and can be made electrically highly conductive too by the introduction of defects. Moreover, ZnO is bio-safe as well as biocompatible and therefore, can be used for biomedical applications without coating.\cite{bio} On the other hand, one of the main limitations of ZnO for its practical applications, is that its band gap is too large to effectively use visible light. Therefore, it is important to have band gap reduction in ZnO. Being a wurtzite crystal structure, one striking feature of the bulk ZnO is the appearance of polar surfaces during its growth along the $c$-axis. The hexagonal wurtzite crystal structure consists of a number of alternating planes composed of tetrahedrally coordinated O$^{2-}$ and Zn$^{2+}$ ions, stacked along the $c$-axis. The oppositely charged ions produce positively charged (0001)-Zn and negatively charged (000$\bar{1}$)-O polar surfaces, resulting in a normal dipole moment and spontaneous polarization along the growth direction.

Nano structured ZnO has received particular attention in current research due to intriguing nanosized effects on their properties. The interesting fact about nano structured ZnO systems, is that they have a diverse group of growth morphologies, such as nanocombs,\cite{comb} nanotubes/nanorods,\cite{tube} nanohelixes/nanosprings,\cite{spring} nanobelts/nanoribbons/nanorings\cite{belt,ring} and nanocages.\cite{cage} The morphology of the ZnO nanostructures is largely directed by the polarity and saturated vapor pressure of the solvents. Recently, doping concentration driven morphological evolution has also been reported for the Fe doped ZnO nanostructure.\cite{Fe-doping} Therefore, the morphology of the ZnO nanostructures can be varied from one shape to another by controlling the growth kinetics through the adjustment of the preparation method and preparation conditions.\cite{shaping}  These various morphologies and matured growth methods in addition with the intrinsic surface and quantum confinement effects of the nano structures, lead to easy preparation of various nano-ZnO based high-tech functional devices. For examples, ZnO nanostructures have been used widely in field-effect transistors,\cite{transistor} light emitters,\cite{light} lasers,\cite{laser} dye sensitized solar cells\cite{dssc} and sensing.\cite{sensor} Nanostructured DMSs have also been of current interest for the future spintronics applications. The possibilities of RT ferromagnetism in ZnO nanostructures, have been realized by the dopings with atoms of TM elements, $sp/d^0$ elements\cite{dzero} or capping the surface of the host system with N or S containing ligands.\cite{capping} There have been many reports which indicate that surface effects and defects, however, play an important role in stabilizing ferromagnetic couplings in these systems.\cite{defect} In fact, the magnetism in the doped ZnO nanosystems is rather subtle and depending on the material preparation conditions, there have been considerable controversy in the reported experimental results of the RT ferromagnetism and its real origin.\cite{dms_contro} It, therefore, clearly indicates the fact that the research field involving the ZnO nanostructures, is still in its infancy but nanostructured materials built up from such nanoparticles are expected to have exceptional properties which are promising in microelectronic devices,\cite{micro} energy conversion and storage,\cite{storage} catalysis\cite{cata} and drag delivery\cite{delivery}. For this reason, it is also essential to identify the exact role of the shape of the ZnO nano particles and to gain a precise understanding of the nature of bonding between the host systems and the added foreign impurities. 

Morphology control of semiconducting nano systems are very demanding for enhancing the light absorption and shortening transfer distance of photo-generated carriers.\cite{nano,zno_nano} Recently, photo-catalytic properties of various ZnO nano structures and the effects of aspect ratio due to the morphological changes, have also been studied.\cite{zno_pec}  In several recent experimental works, the morphology-controlled synthesis of ZnO nanostructures has been demonstrated in terms of the interplay of several associated effects like polar charges, surface area, elastic deformation and so on\cite{nmat_orientation,expt_shape_control}The variations of the dimensionality for the ZnO nanostructures, on top of their finite size effects, incur several unique properties. Therefore, comparison of their properties with respect to that of the well-studied bulk, surface and thin-film systems related to ZnO, would be useful as well as interesting. Theoretical study on the magnetism and band gap variation of the ZnO nanosystems with shape changes is, however, very limited. One possible reason may be the absence of any well defined recipe to construct nano crystals with a defined aspect ratio, apart from a few exceptions.\cite{ito,sdatta} 

In the present work, we report our results on first principles electronic structure calculations to investigate the morphology effects of host system on the possibilities of engineering the magnetic couplings and band gap narrowing of the ZnO nano structures upon diluted substitutional doping of the two 3$d$ late TM atoms of same type at the two metal sites of the host systems. To explore the morphology effects of a (ZnO)$_{24}$ nano system, we have considered its three structural forms of different shapes - a 2D mono-layer-type nanosheet, an 1D nanotube and one finite quasi-spherical cage-shaped nanocluster, which have the same number of atoms with stoichiometric composition. The three structures of the host system are, rather, distinguished by different degrees of spatial confinement. We study the substitutional bi-doping with each of the 3$d$ late transition metal atoms Mn, Fe, Co, Ni and Cu at the metal sites of the (ZnO)$_{24}$ nano system for all the three structures of the host system. We have employed sophisticated hybrid functional as prescribed by Heyd, Scuseria and Ernzerhof (HSE)\cite{hse} for accurate description of the electronic structure. Our study reveals that the effect of shape change is dramatic in case of the band gap variation of both the pristine as well as the doped systems in going from one morphology to the others. The effect of this morphology change on the magnetic properties of the doped systems, is, however, comparatively minor as the magnetic coupling remains mostly anti-ferromagnetic irrespective of the morphology changes and the notable difference arises mainly in the nature of the spatial separation of the two dopant impurity atoms in cases of the most favorable bi-doped systems. We believe, our results here will not only be useful to provide a basis to rationalize the diverse experimental reports in the past on the TM-doped semiconductor nano particles, but may also guide the future experiments to opt a specific synthesis method for the desired opto-electronic properties.

\begin{figure*}
\rotatebox{0}{\includegraphics[height=5.0cm,keepaspectratio]{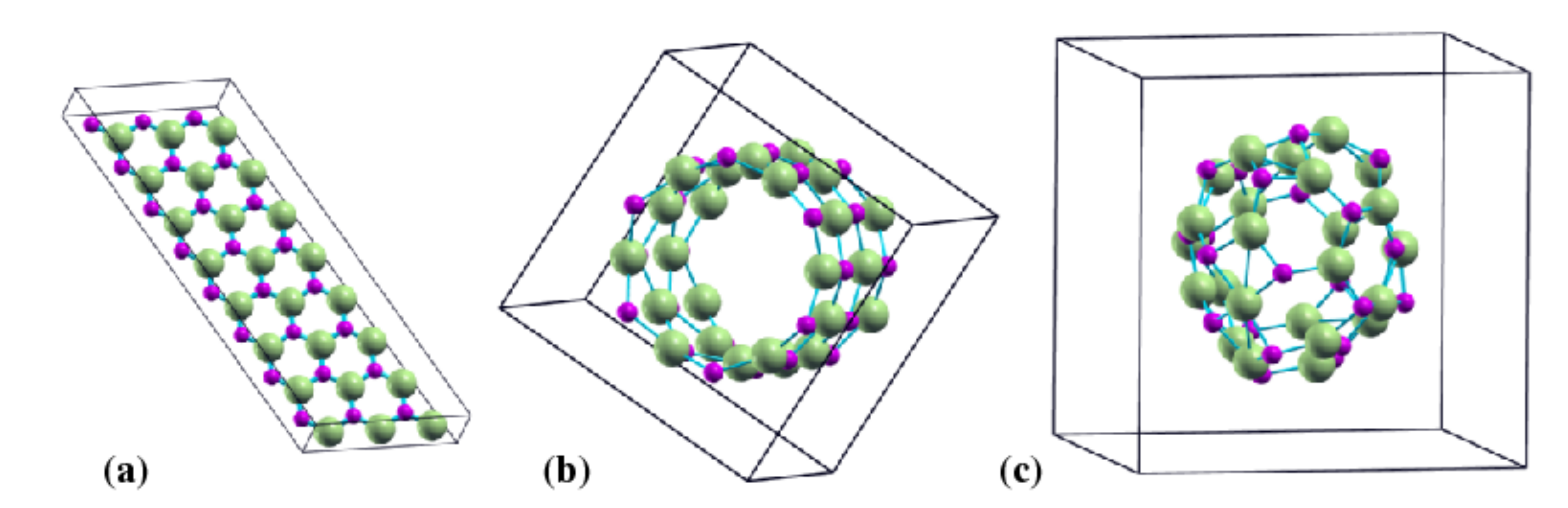}}
\caption{(Color online) Three different morphologies of (ZnO)$_{24}$ nano structures used in this work : (a) 2D extended sheet, (b)1D elongated tube and (c) cage-like ZnO cluster of finite dimension. Green colored larger balls correspond to Zn atoms and violet colored smaller balls correspond to O atoms.}
\label{mes}
\end{figure*}

\section{\label{methodology} Computational Details} The calculations reported in this study, were performed using DFT within the framework of pseudo potential plane wave method, as implemented in the Vienna abinitio Simulation Package (VASP).\cite{kresse2} 3$p$, 3$d$ as well as 4$s$ electrons for the transition metal atoms and 2$s$ as well as 2$p$ electrons for the oxygen atoms, were treated as valance electrons. The interaction between the valance electron and ion core has been described by the projected augmented wave (PAW) pseudo potential \cite{blochl,kresse} and exchange correlation energy density functional was considered under the framework of generalized gradient approximation (GGA) as formulated by Perdew, Burke and Ernzerhof (PBE).\cite{perdew} Energy cut-off of plane wave expansion was set to 450 eV. The convergence of the energies with respect to the cut-off value were checked. Symmetry unrestricted geometry optimizations were performed using the conjugate gradient and the quasi-Newtonian methods until all the force components were less than a threshold value of 0.001 eV/{\AA}. The convergence of self consistency was achieved with a tolerance in total energy of 10$^{-4}$ eV. It is well known that calculations with the GGA-PBE level of theory generally underestimates the band gap and formation energy of transition metal oxide semiconductors.\cite{dft_drawback} To overcome these deficiencies, all electronic property calculations have been carried out using HSE hybrid density functional,\cite{hse} which mixes 25 $\%$ of the exact nonlocal exchange of Hartree-Fock theory with the semilocal PBE functional. The GGA-PBE relaxed structures were used as input structures for the HSE calculations. Each of the three structures, has been modeled with a periodic super-cell. A rectangular parallelepiped super-cell for the nanosheet geometry, a square parallelepiped super-cell for the nanotube geometry and a simple cubic super-cell for the nanocage geometry were used with periodic boundary condition. The vacuum size was set at 12 {\AA}  between the system and its neighboring images, which essentially makes the interaction between them negligible. The vacuum separation was considered along the vertical direction for the nanosheet structures, along the cross-sectional direction for the nanotube structures and along the all directions around each nanocage structure. The Brillouin zone has been sampled with a sufficient number of $\Gamma$-centered k-point sets generated based on the Monkhorst and Pack scheme.\cite{horse} Full relaxation of the geometries have been done with k-point mesh of 4$\times$4$\times$1 for the nanosheet geometry, 4$\times$1$\times$1 for the nanotube geometry. For the nanocage geometry, the reciprocal space integrations were carried out at the $\Gamma$ point. Spin-polarized calculations were performed for all the transition metal doped systems.

\section{\label{results}Results and Discussions}

\subsection{\label{pure}Pure nano-structures}

Fig. \ref{mes} shows the optimized minimum energy structure (MES) for each of the three morphologies of the pristine (ZnO)$_{24}$ system. The optimized structure of the (ZnO)$_{24}$ nano-sheet having a length of 26 {\AA} and a width of 9.8 {\AA} consists of hexagonal (ZnO)$_3$ building blocks  with the infinite periodicity all along the plane. The parallelogram shape for the planar sheet has been considered to have the provision of including both the {\it near} and {\it far} spatial separations between the two dopant atoms in the case of bi-dopings in this structure. Our calculated average Zn-O bond length for the optimal nanosheet is 1.88 {\AA} in accordance with the earlier result,\cite{ml} while its value for the bulk wurtzite ZnO is 1.98 {\AA} corresponding to its two lattice constants of {\it a} = 3.249 {\AA} and {\it c} = 5.207 {\AA}.\cite{lat} In the optimized nanosheet structure, the average values of the Zn-O-Zn and O-Zn-O angles remain at 120$\degree$. The (ZnO)$_{24}$ nanotube has a regular zigzag (6,0) form and again consists of (ZnO)$_3$ hexagons. The optimized structure of the nanotube has a length of around 11 {\AA} and diameter of 7 {\AA}. The average Zn-O bond-length of the optimized nanotube structure, is 1.91 {\AA}. The Zn-O-Zn bond angles along the axial direction are around 109.3$\degree$-109.5$\degree$ and along the cross sectional direction are around 119.2$\degree$-119.8$\degree$. The corresponding values for the O-Zn-O bond angles are around 112.5$\degree$-112.7$\degree$ along the axial direction and around 133.2$\degree$-133.6$\degree$ along the cross sectional direction. The optimal structure of the fullerene-like (ZnO)$_{24}$ nanocage, on the other hand, consists of several sections of (ZnO)$_4$ octagons, (ZnO)$_3$ hexagons and (ZnO)$_2$ squares as reported in earlier works.\cite{cage2,cage3} The appearance of the octagon, hexagon and square shaped building blocks together in case of the optimized (ZnO)$_{24}$ nanocage structures, arises in order that Zn-Zn or O-O direct bonds are not formed.  The diameter of the optimal pristine cage-shaped structure is $\approx$ 11 {\AA} and the average value of Zn-O bond-lengths  is 1.94 {\AA}, which is slightly larger than that of the pristine nanosheet as well as the nanotube structures. It is to be noted that recently, the planar structure of a flat honeycomb nanosheet\cite{flatsheet} and formation of the smooth single-walled ZnO nanotube\cite{zno_tube} as well as the cage-shaped structure for larger (ZnO)$_n$ clusters with $n>$8,\cite{cage3} have also been predicted to be energetically more favorable than their reconstructed geometries.

\begin{figure}
\rotatebox{0}{\includegraphics[height=8.8cm,keepaspectratio]{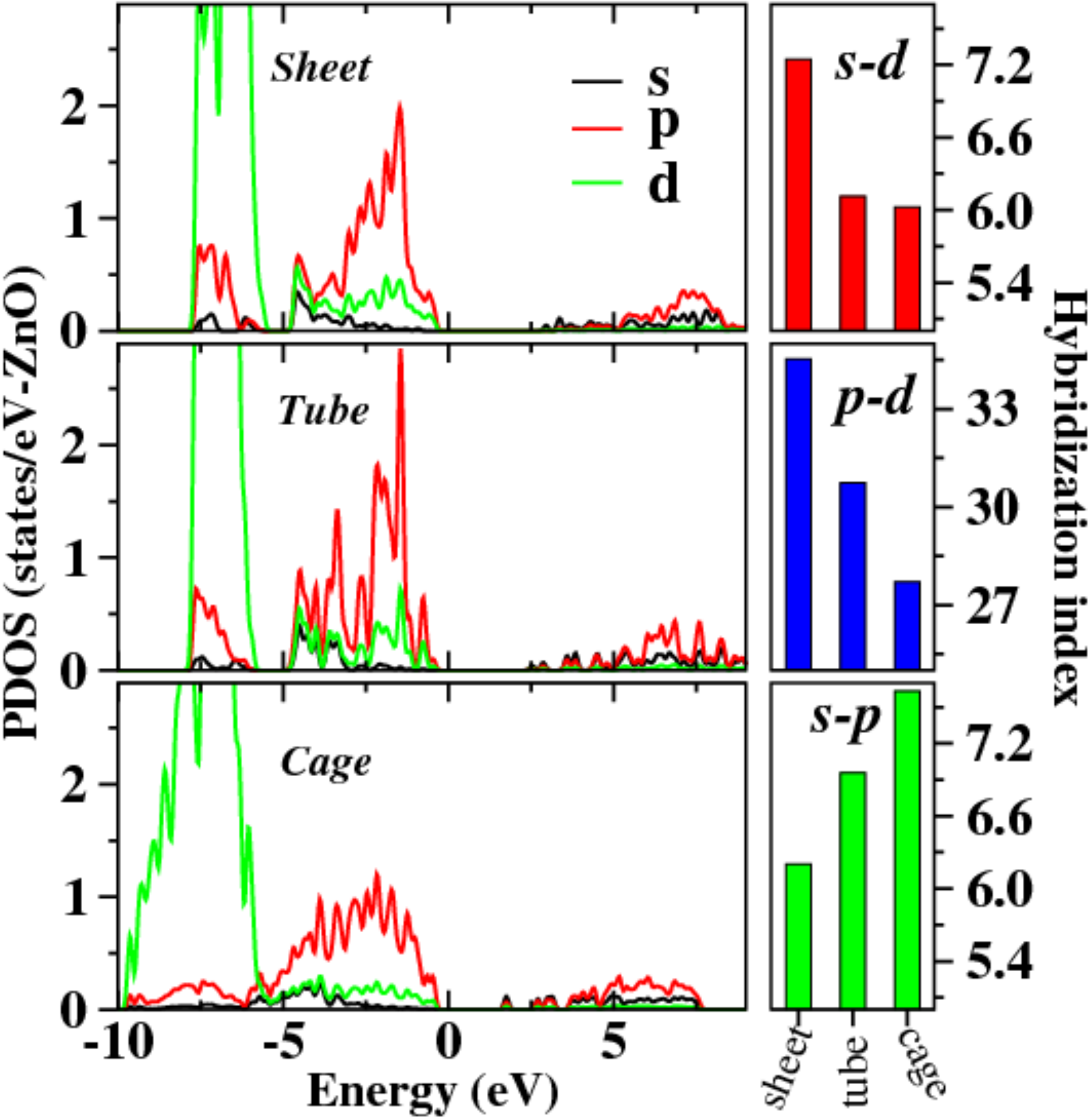}}
\caption{(Color online)Plot of HSE method calculated $s$, $p$ and $d$ orbital projected density of states (for the nanosheet, nanotube and cage shaped structures in the left panels) and $s$-$d$, $p$-$d$ as well as $s$-$p$ hybridization indexes (from top to bottom in the right panels) corresponding to the MESs of the three pure (ZnO)$_{24}$ nano structures. A smearing width of 0.1 eV is used. The energy along x-axis of the left panels, is shifted with respective to the Fermi energy of the respective system.}
\label{dos_pure}
\end{figure}

The different geometrical structures of the three systems, are primarily characterized by different aspect ratio ({\it i.e} length/width ratio) which is decreasing along nanosheet$\rightarrow$nanotube$\rightarrow$nanocage. It would be interesting to distinguish them in terms of their microscopic electronic properties. Fig. \ref{dos_pure} shows the $s$, $p$ and $d$ orbitals projected density of states (PDOS) summed over all the constituent atoms for the optimal structures of the three morphologies in case of the pristine (ZnO)$_{24}$ nano system. It is seen that the top of the valence band (TVB) is mostly contributed by the Zn$-d$ and O$-p$ orbitals. On the other hand, the bottom of the conduction band (BCB) is mainly contributed by the O$-p$ orbitals and also partially by the O$-s$ and Zn$-s$ orbitals. We note that the change in morphology affects the types of orbital hybridization for the constituent atoms of the (ZnO)$_{24}$ system. We find out that the optimal structure for each of the three morphologies of the pristine (ZnO)$_{24}$ nano structure, is characterized by the dominance of the different kinds of orbital hybridization. In order to understand the trend in the  hybridization among the various orbitals of the constituents in the optimal structures, we have analyzed a hybridization index parameter as shown in the right panels of the Fig. \ref{dos_pure}. We define the hybridization index parameter\cite{hdi1,hdi2} h$_{kl}$ as, $h_{kl}=\sum\limits_{I=1}^{40}\sum\limits_{i=1}^{occ}w_{i,k}^{(I)}w_{i,l}^{(I)}$ ; where $k$ and $l$ are the orbital indices - $s$, $p$ and $d$, $w_{i,k}^{I}$ ($w_{i,l}^{I}$) is the projection of $i$-th Kohn-Sham orbital onto the $k$ ($l$) spherical harmonic centered at atom $I$, integrated over a sphere of specified radius. First of all, it is seen from the Fig. \ref{dos_pure} that the $p$-$d$ hybridization index is the most dominating for each of the three pristine structures, with the maximum value for the optimal (ZnO)$_{24}$ nanosheet structure. In addition, the optimal (ZnO)$_{24}$ nanosheet is associated with the maximum value of the $s$-$d$ hybridization index too, which favors a planar geometry. The (ZnO)$_{24}$ nano-cage, on the other hand, is associated with the enhanced $s$-$p$ hybridization index which shows a decreasing trend along cage$\rightarrow$tube$\rightarrow$sheet structures. The optimized nanotube structure, however, always adopts the intermediate value for each of the three hybridization index parameters.

The optimized geometries of the three pristine (ZnO)$_{24}$ nano structures are found to be nonmagnetic. Our calculated band gaps [{\it i.e.} HOMO-LUMO gap] for the pristine systems, show a nice trend with the change of aspect ratio. The calculated band gaps using the HSE functional, are 3.39 eV, 3.0 eV and 2.22 eV for the optimal nanosheet, nanotube and the cage-shaped (ZnO)$_{24}$ nano structures respectively. While the band gaps of the tube and cage structures are dependent on their respective diameters,\cite{orientation} our calculated band gap for the ZnO monolayer is in close agreement with the earlier result of 3.57 eV by GW calculations.\cite{bg_bulk_theory} It is also important to note that our calculated band gaps are of direct type for all the three pristine structures, in accordance with the direct wide-band gap of 3.3 eV for the bulk wurtzite ZnO.\cite{bg_bulk} Interestingly, this fact of possessing same kinds of band gaps for the bulk as well as monolayer phases of the ZnO system, is in contrast with the cases of transition metal dichalcogenides, namely MoX$_2$, WX$_2$ (X = S, Se, Te) which are indirect band gap semiconductors in bulk phase, whereas their monolayers have direct band gaps.\cite{chalco1,chalco2} The trend of the decreasing band gaps with the decreasing aspect ratios, as mentioned above for the (ZnO)$_{24}$ nano systems, is, however, in accordance with the trend in band gap variation of another II-VI semiconductor, namely CdS nanocrystals with the change of aspect ratio.\cite{bg_pure} We note that the decrease in band gap of the pristine (ZnO)$_{24}$ nano structures in going from the nanosheet to the cage-shaped geometry, arises mainly from the downward shifting of the BCB with the decreasing aspect ratio.

\subsection{\label{doped}Bi-doped nano-structures}
The bi-dopings with the TM atoms tunes the band structure of the host systems such that it promotes band gap engineering, in addition to inducing the spin polarization in the doped systems. In this section, we have analyzed the magnetic properties as well as variation of the band gaps of the doped systems and ultimately figured
 out the effects of the morphological changes of the host system with decreasing aspect ratio from the nanosheet to the nanocage geometry.

\subsubsection{\label{mag}Understanding the magnetic interactions}
\begin{figure}
\rotatebox{0}{\includegraphics[height=7.8cm,keepaspectratio]{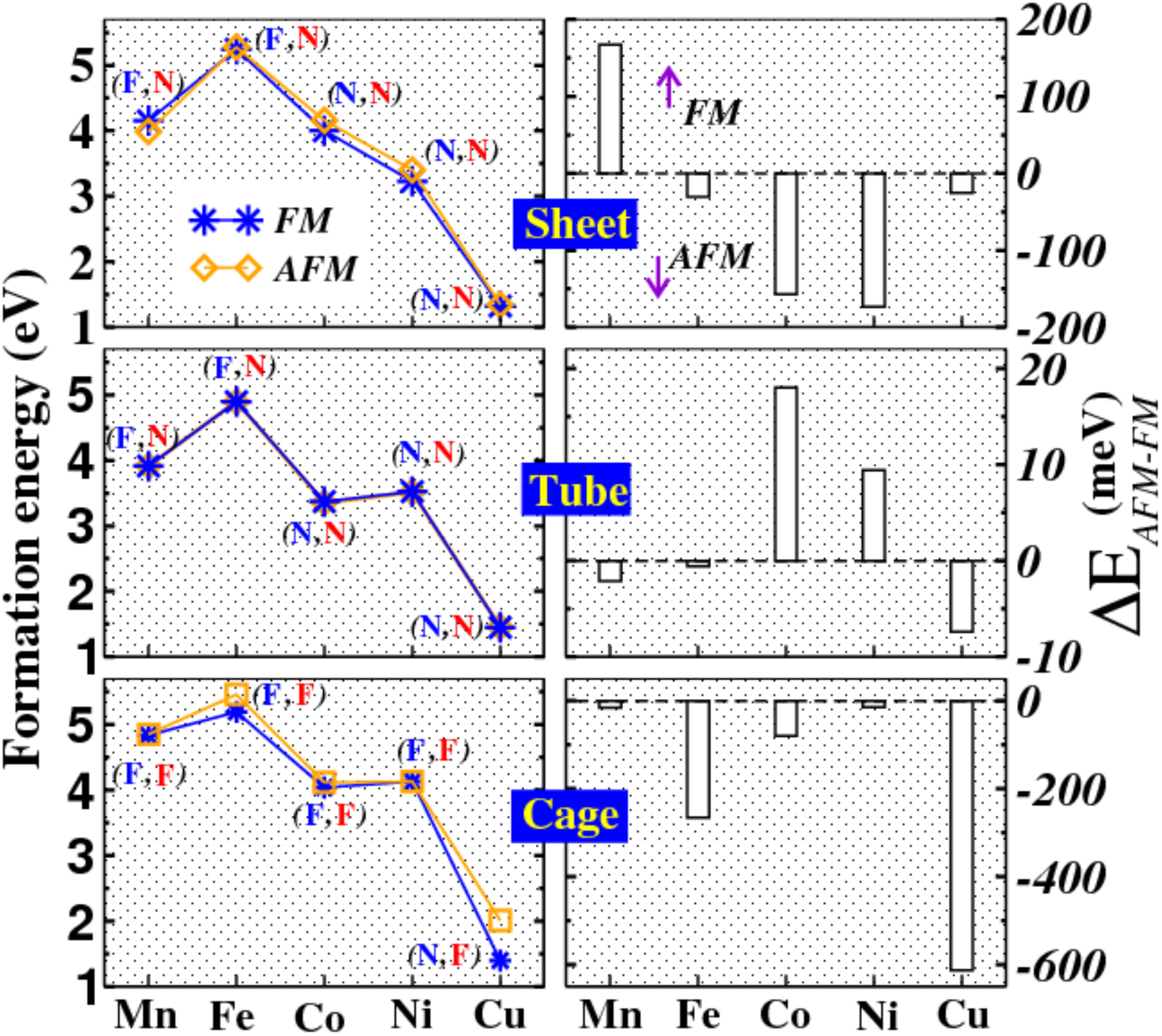}}
\caption{(Color online) Plot of formation energies as calculated by HSE method for the optimal ferromagnetic (FM) and anti-ferromagnetic (AFM) coupling configurations of the dopants elements (left panels) and energy difference, $\triangle E$ between the optimal AFM and FM structures for a given dopant (right panels) in case of bi-doped (ZnO)$_{24}$ nano structures. The symbols within parenthesis, denote the types of spatial separation between the two dopants for the optimal FM as well as AFM couplings.}
\label{magnetic}
\end{figure}

The stability of TM bi-doped ZnO systems, is found to be sensitive to the magnetic coupling or chemical bonding between the dopant atoms,\cite{adhesion}which in turn is expected to be influenced by the change in morphology of the host systems because of the different degrees of spatial confinement of the various morphologies. In order to explore the effects of the change in morphology of the host system on the magnetic interactions between the two magnetic dopant atoms, we have first examined the preferred magnetic orderings of a pair of each magnetic atom M = Mn, Fe, Co, Ni and Cu in the three morphologies of (ZnO)$_{24}$ nano system and thereafter, have tried  to understand its origin. In a recent work on Mn-doped ZnO nano clusters,\cite{substitute} it was shown that substitutional doping at the Zn site is energetically more favorable compared to endohedral or exohedral doping. Therefore, we have considered in our study here only the substitutional doping with the TM atoms. The substitution at two Zn sites of the host system by the two dopant M atoms, thereby corresponds to a doping concentration of 8.4 atom $\%$. To determine the energetically most favorable magnetic coupling, we have performed calculations for both the ferromagnetic as well as anti-ferromagnetic couplings of the two dopants. Moreover, for each of the ferromagnetic and anti-ferromagnetic orderings, both the {\it near} as well as {\it far} spatial separations of the two dopant atoms have been explored. In case of the {\it near} spatial separation, the two TM dopant atoms are separated by one oxygen atom, while for the case of {\it far} separation, we consider as much separation as possible in the chosen geometries. It is important to note here that the solubility of the dopant atoms in the (ZnO)$_{24}$ host systems at the given growth conditions, can be determined by calculating formation energy for the bi-doped systems. The formation energy is the energy needed to insert the two dopant atoms (taken from a reservoir) into the (ZnO)$_{24}$ systems after removing two Zn atoms from the host system (to a reservoir). We, therefore, define the formation energy for the two $M$ atoms substitutional doping at two Zn sites, as $\triangle E_f = E(Zn_{24}O_{24}) - E(Zn_{22}O_{24}M_2) -2E(Zn)+2E(M)$, where  E(Zn$_{22}$O$_{24}$M$_2$) is the total energy of the optimal bi-doped system and E(M) [E(Zn)] is the total energy of an isolated M atom [ Zn atom ] which resembles the chemical potential of the respective elements in case of bulk-like system. According to our definition, formation energy should be positive for a stable structure. Generally, high formation energy leads to a low solubility. Note that the values of the single atom energies in the expression of the formation energy, depends very much on the growth conditions under which the bi-doped systems are prepared. In our calculations of formation energies, we have taken the isolated atoms in gas phase, as the reservoir of the respective element. The left panels of the Fig. \ref{magnetic} show the plot of our calculated formation energies for the optimal ferromagnetic as well as optimal anti-ferromagnetic configurations. The preferred spatial separations between the two impurity dopant atoms in case of the optimal FM as well as AFM orderings, are also conveniently mentioned in the Fig. \ref{magnetic} by a notation like `({\it1st}, {\it2nd})'. The notation {\it 1st} denotes the type of favorable spatial separation for the optimal FM ordering out of {\it far} (denoted by F) as well as {\it near} (denoted by N) spatial separations and the notation {\it 2nd} denotes the type of favorable spatial separation for the optimal AFM ordering. Therefore, each of the {\it 1st} and {\it 2nd} notations, can adopt either `F' or `N' symbol. It is seen that the bi-Cu doping has the least formation energy and therefore, it will be the most favorable case than the other bi-dopings for each of the three host systems. We have also performed the formation energy calculations with respect to the crystalline bulk phases of the metal atoms i.e $\alpha$-Mn, body centered cubic Fe, hexagonal closed pack Co, face-centered cubic (FCC) Ni, FCC Cu and hexagonal closed pack Zn structures as the reservoir of the respective element. We find that the overall trend in the variation of the formation energies remains the same in the both cases i.e decreasing formation energy for the dopants with increasing atomic numbers. For the Cu bi-doping in each of the three morphologies in particular, the formation energy, however, becomes negative with respect to the reservoir of Cu in equilibrium with the bulk FCC Cu. Note that the negative formation energy (according to our definition) of the Cu bi-doped  (ZnO)$_{24}$ system, implies relative difficulty for fabrication of this system in experiment as additional energy is needed.

Further, we use the energy difference $\triangle$E between the optimal AFM and FM orderings (${\it i.e}$ $\triangle E$ = E$_{AFM}$-E$_{FM}$) as an indicator of the magnetic stability. A positive (negative) $\triangle E$ indicates that the ground state is FM (AFM). Our estimated values of $\triangle E$ for each doped system, are also shown in the right panels of the Fig. \ref{magnetic}. It is, therefore, seen that overall the AFM coupling between the two dopant atoms is energetically more favorable, except the case of bi-Mn doping in case of the 2D nanosheet geometry of the (ZnO)$_{24}$ nano structure. It is also interesting to note that the optimal AFM couplings in the sheet geometry, always favor a {\it near} spatial separation of around 3.14 - 3.26 {\AA} between the two dopant atoms for all the bi-doping cases. In case of the most stable structure  for the bi-Mn doping in the (ZnO)$_{24}$ nano sheet structure, the two Mn atoms favor to couple ferromagnetically with a {\it far} spatial separation of 6.52 {\AA}. This is also indicating a fact that the a homogeneous distribution of Mn atoms in the ZnO nanosheet structure, would favor ferromagnetism. Note that our observation for the magnetic couplings of the bi-doping cases in the nanosheet structure, is in accordance with the previous results on dopings in a ZnO thin film, which indicated FM coupling for Mn-doped ZnO thin film and the AFM coupling for each of the Fe, Co, Ni doped ZnO thin films.\cite{thinfilm1,thinfilm2}In the case of the Mn-doped ZnO thin film, the magnetic coupling has, however, been reported to depend very much on the Mn doping concentrations. For the cases of our tube-shaped (ZnO)$_{24}$ nano structure, the energy difference, $\triangle E$ is found to be very small of the order of the RT energy ({\it i.e.} 25 meV), indicating a borderline case which means both the FM and AFM couplings are almost degenerate. Above all, the properties of a nanotube structure, are very much dependent on the size and aspect ratio of the tube.\cite{shape_dependent1,shape_dependent2} However, it is not our purpose to explore it here as we are focusing on the overall variation in the properties with the change of aspect ratio. Finally, the trend for the TM bi-doping in case of the cage-shaped (ZnO)$_{24}$ nano structure is, however, solely of AFM type including the case for the bi-Mn doping. It is also interesting to note that the most stable AFM coupling for each bi-doping case in the cage structure, always adopts a {\it far} spatial separation between the two dopants. Note that our findings of favorable magnetic interactions of the bi-dopings in the cage structure, are also in accordance with the available earlier reports from the literature. Liu {\it et al.}\cite{liu} showed that doping of Mn into the Zn$_{12}$O$_{12}$ cluster, stabilizes AFM ground state for small Mn-Mn separation, while FM and AFM states are degenerate for large Mn-Mn distance. Another work on Fe doped ZnO nano clusters, indicates a clean dominance of AFM coupling in a neutral defect-free cluster, whereas defects under suitable conditions, can induce FM interaction between the dopant atoms.\cite{ganguli} To point out precisely the overall effects of the change of morphology on the magnetic interactions, it is seen that the magnetic coupling in the most stable structure of the Mn bi-doped systems, changes from {\it far} FM coupling in case of the sheet geometry to {\it far} AFM coupling in case of the cage shaped geometry. For the other bi-doping cases, the ground state structures favor AFM coupling for both the sheet as well as cage shaped geometries of the host system. The only difference in case of the magnetic couplings for the MESs of the bi-dopings with the Fe, Co, Ni and Cu ions between the sheet and cage structures of the host system, arises in the spatial separation of the two dopant atoms. This is in the sense that the AFM coupling of the two dopants at a {\it near} spatial separation is transformed to an AFM coupling at {\it far} spatial separations while transforming the shape of the host system from the sheet to the cage geometry. It is also interesting to note that the value of energy difference, $\triangle E$ is exceptionally larger in case of the bi-Cu doping in the cage-shaped structure compared to its value in case of the sheet morphology.

\begin{figure}
\rotatebox{0}{\includegraphics[height=7.8cm,keepaspectratio]{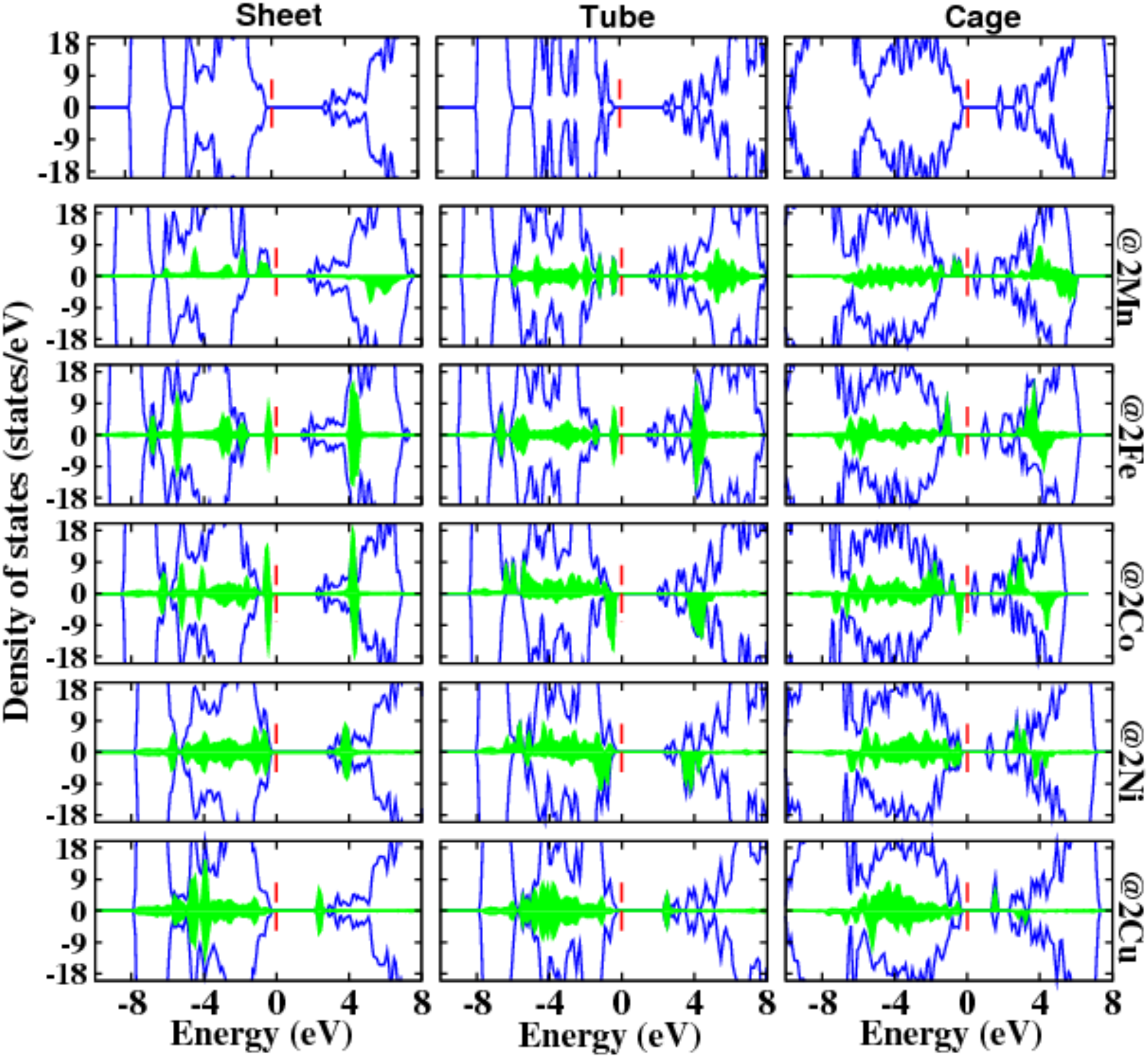}}
\caption{(Color online) Plot of TDOS (blue line) and PDOS (shaded region) of the most stable structure for the TM bi-doped (ZnO)$_{24}$ nanostructures in each of the three morphologies calculated within HSE method. The green colored shaded region shows the TM-$d$ PDOS. For clarity, the TM-$d$ PDOS is scaled by a factor of two for each system. TDOS of the pristine systems are also shown at the top panels. A smearing width of 0.1 eV is used. The energy along x-axis is shifted with respective to the Fermi energy of the respective system, as indicated by a vertical dashed line passing through zero energy. The positive and negative values of DOS are for the majority and minority spin channels, respectively. }
\label{pdos}
\end{figure}

  While studying the local structure around the TM dopants in the MESs of the doped systems, we note that the $\langle$TM-O$\rangle$ bond length with the neighboring oxygen atoms, is slightly larger than the $\langle$Zn-O$\rangle$ bond-length of the corresponding pure host system in case of the Mn bi-doping, while it is slightly smaller in case of the bidopings of Fe, Co, Ni and Cu atoms. Overall, the  $\langle$TM-O$\rangle$ bond lengths are larger in the doped nanocage structure compared to that in the doped nanosheet structure for each TM bi-doping. The significant overlaps of the valence $d$-orbital of the dopant atoms with the $p$-orbital of  its neighboring oxygen atoms in the doped (ZnO)$_{24}$ system, results into a change in  the electronic configurations of the interacting atoms and thereby induces a spin polarization to them. In order to understand the magnetic properties of the bi-doped systems in more details, we have plotted in Fig. \ref{pdos} the spin-polarized total density of states (TDOS) and the dopants TM-$d$ PDOS for the most stable structure of each doped system. It is seen that the hybridization between each dopant atom and its neighboring oxygen atoms, results in the splitting of the energy levels near the Fermi energy. In case of the MES of bi-Mn doping in the sheet structure, the majority spin channel is completely filled and the minority spin channel is completely empty. For the MESs of all the other bi-doping cases in the sheet as well cage structures, the TDOS is symmetric around the Fermi energy, which is a typical signature of the AFM coupling. The top of the valence band of the doped systems, is contributed mainly by the dopant states and therefore, the bidopings also play role in band gap engineering as discussed in the following Section \ref{gap}.

\begin{figure}
\rotatebox{0}{\includegraphics[height=9.5cm,keepaspectratio]{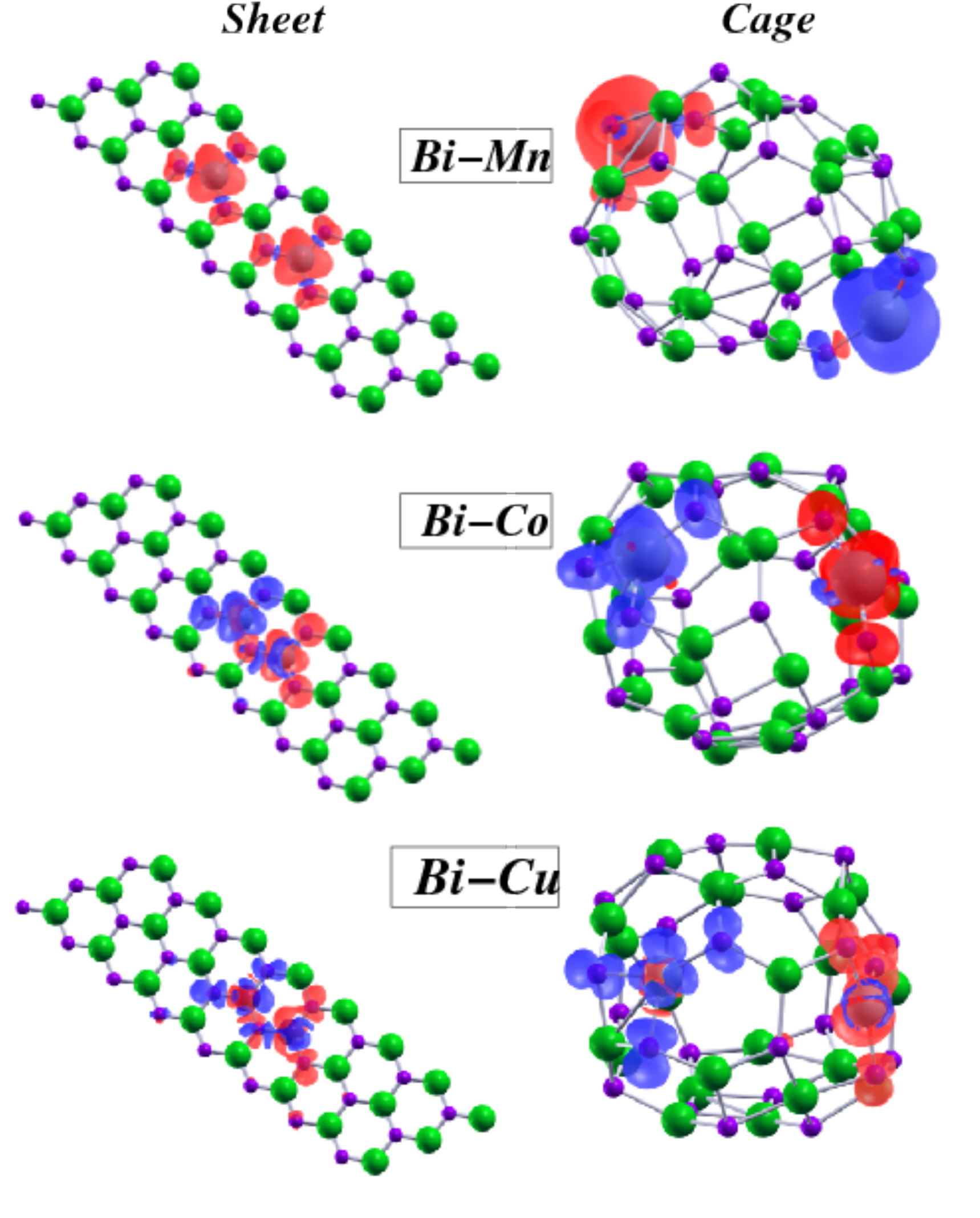}}
\caption{(Color online) Spin density surface plot of the most stable structures for the Mn, Co and Cu bi-doped (ZnO)$_{24}$ nanostructures in the nanosheet (left) and nanocage (right) morphologies calculated within HSE method. The isosurface value of spin charge is 0.01 e$^-$/{\AA}$^3$. The red (blue) colored surface indicates the positive (negative) spin. }
\label{spin}
\end{figure}

 Mulliken population analysis\cite{muli} of spin has been carried out to calculate the atom-centered magnetic moments of the constituent atoms in the MES of all the doped systems for the three morphologies. Fig. \ref{spin} shows the spin density surface plot for the MESs of three representative bi-doped systems, namely Mn, Co and Cu bi-doped systems in the sheet and cage morphologies. In the most optimized structures for the bidopings in the sheet geometry, our calculations show that the TM-TM distances vary as 6.52 {\AA}, 3.18 {\AA}, 3.24 {\AA}, 2.89 {\AA} and 3.12 {\AA} for TM = Mn, Fe, Co, Ni and Cu respectively. We note that the two dopant atoms carry an average local magnetic moment of magnitude 4.45 $\mu_B$, 3.61 $\mu_B$, 2.65 $\mu_B$, 1.66 $\mu_B$ and 0.74 $\mu_B$ per dopant site for Mn, Fe, Co, Ni and Cu respectively. The moments of the neighboring O atoms are, however, very small within an amount of 0.04-0.07 $\mu_B$ per oxygen atom. In the specific case of the FM ground state for the bi-Mn doping in the (ZnO)$_{24}$ nano sheet structure, we note that the nearest neighbor oxygen atoms around the TM-dopant atoms, carry an average moment of 0.04 $\mu_B$ per O-site and they are also ferromagnetically coupled with the two dopant atoms. Therefore, two Mn atoms are interacting via ferromagnetically coupling to the oxygen atoms. The ferromagnetic coupling for Mn-doped DMS systems, has been understood previously in terms of hole mediated RKKY interaction or strong $p$-$d$ indirect interaction for both the bulk as well as nano systems.\cite{rkky,pdindirect} On the other hand, in case of the AFM ground states for the bidopings with Fe, Co, Ni and Cu in the sheet structure with near spatial separation, we note that the two dopant atoms are connected to each other through one common oxygen atom which has zero magnetic moment {\it i.e} this oxygen atom is behaving as a nonmagnetic atom. Therefore, the two dopant atoms and the moments associated with the other neighboring oxygen atoms, interact with each other through a {\it superexchange} interaction involving the intermediate nonmagnetic oxygen atom, which results in AFM coupling within each other. In case of the bi-dopings in the nanotube structure of (ZnO)$_{24}$ systems, the values of the local magnetic moments for the dopant atoms remain almost the same as that of the respective cases of the bi-doped nanosheet structure. Turning our attention to the MESs of the bi-doped cage structures, we note that the TM-TM separations are 8.9 {\AA}, 8.4 {\AA}, 5.9 {\AA}, 6.1 {\AA} and 5.7 {\AA} for TM = Mn, Fe, Co, Ni and Cu dopings, while the average local magnetic moments at the dopant site retain values of 4.42 $\mu_B$, 3.60 $\mu_B$, 2.65 $\mu_B$, 1.65 $\mu_B$ and 0.68 $\mu_B$ respectively. The favorable magnetic coupling between the two dopant atoms, is still of AFM type as it was for the MESs in case of the most bi-doped nanosheet structures. The striking point to note is that contrary to the bi-doped nanosheet structures, the two dopant atoms in the MESs of the bi-doped nanocage structures, like to stay away from each other, which signifies a long range nature of the magnetic interaction and it may be regarded as a possible consequence of the finite size effect  of the cage morphology.

\begin{figure}
\rotatebox{0}{\includegraphics[height=6.5cm,keepaspectratio]{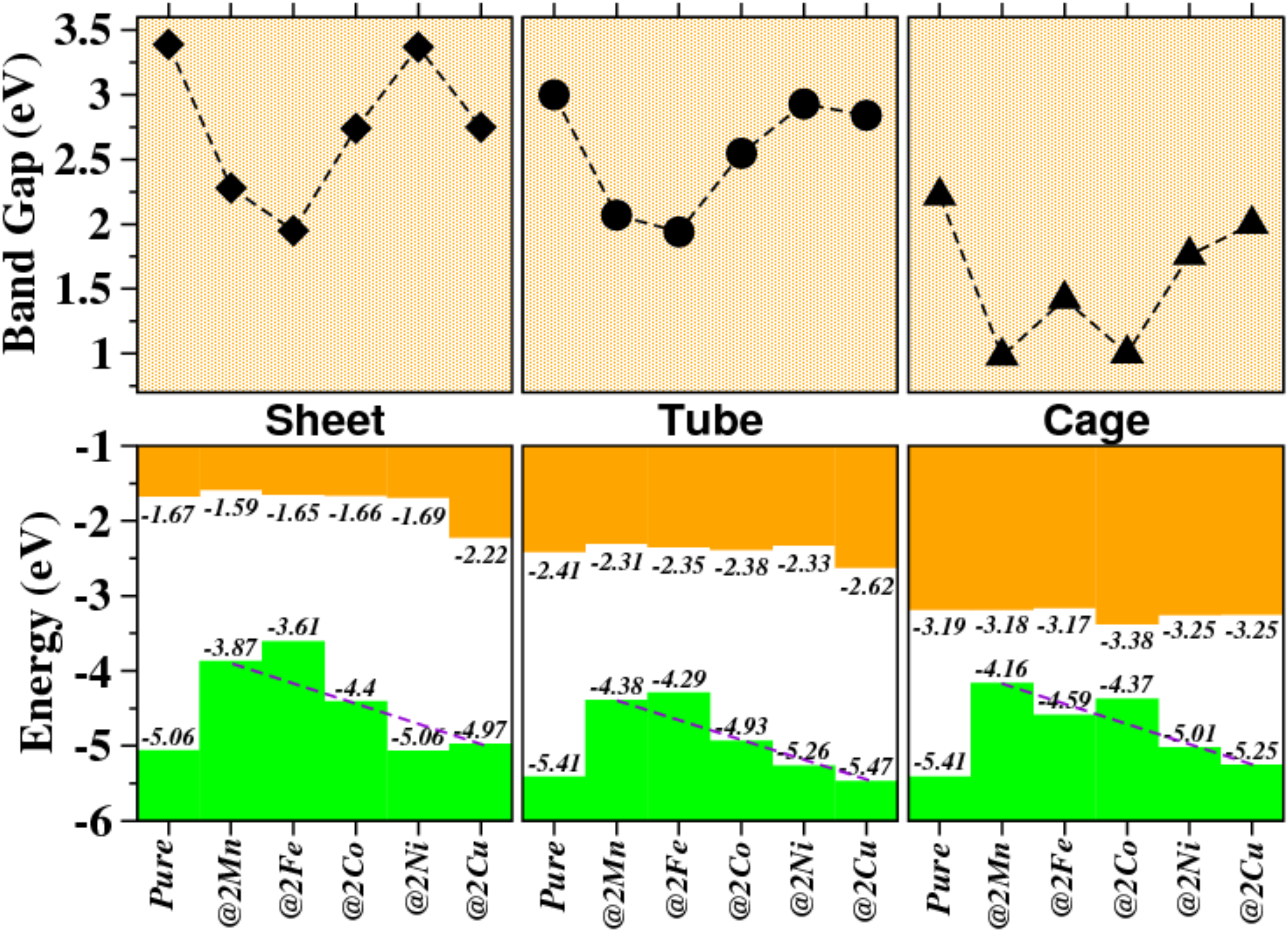}}
\caption{(Color online) Plot of calculated band gaps (upper panels) and band-edge energies (lower panels) by HSE methods for the MESs of the pure as well as TM bi-doped (ZnO)$_{24}$ nanostructures in the three morphologies. The green (orange) colored shaded regions in the bottom panels represent the valence (conduction) energy region. The dashed lines in the lower panels show the overall trend in the band gap variation of the bi-doped systems.}
\label{offset}
\end{figure}

\subsubsection{\label{gap}Understanding the trend in band gap variations}
Our calculated band gap [the highest occupied molecular orbital - lowest unoccupied molecular orbital (HOMO-LUMO) gap] using HSE functional for the MESs of the pristine as well as the doped systems of all the three structures, are shown in Fig. \ref{offset}. The positions of the TVB and BCB are also shown in the Fig. \ref{offset}. We find that the band gaps in general reduce with the decreasing aspect ratio of the host systems {\it i.e} changing the morphology from the nanosheet to the nanocage structures for both the pristine as well as the doped systems. This means that a bi-doped system, in general, possesses the maximum band gap in the sheet geometry and the least for the cage-shaped geometry. This reduction in band gaps of the doped systems arises mainly from the downward shifting of the BCB of the system while moving along the nanosheet$\rightarrow$nanotube$\rightarrow$nanocage structure, as clearly seen from the band edge positions of each doped system in the lower panels of the Fig. \ref{offset}. Also note that the band gaps for the doped systems are, in general, reduced compared to that of the pristine system within a given morphology having fixed degree of spatial confinement. It can be readily understood from the PDOS plot in the Fig. \ref{pdos} which shows the appearance of the dopant states in the mid gap region of the pristine (ZnO)$_{24}$ system and thereby resulting into the lowering of effective band gaps of the doped systems. Interestingly, the band gaps of the doped systems within a given morphology of the host system, show an overall increasing trend from the bi-Mn doped system to the bi-Cu doped system. It can be understood from the dopant's electronic configuration. In case of bi-Mn doping, each dopant atom is mainly characterized by half-filled $d$-orbital. While considering the bi-doping cases along Mn$\rightarrow$Fe$\rightarrow$Co$\rightarrow$Ni$\rightarrow$Cu, the electronic states of each dopant atom is gradually filled by one extra electron. This increases the splitting of the $d$-orbital of the dopant atoms and shifts the TVB towards lower energy. Consequently, the TVB gradually moves downward along bi-Mn doped system towards bi-Cu doped system as shown by the dashed lines in the Fig. \ref{offset} for the three structures. It is, therefore, obviously seen that the calculated band gaps of the doped systems, show a wide variation with the change of morphology of the host systems, as both the effects of changing morphology as well as the electronic structure of the dopant atoms come into play.

\section{Summery and Conclusions}
In summary, first principles electronic structure calculations have been performed to investigate the effects of morphological changes of the (ZnO)$_{24}$ host system, on the magnetic properties and band gap variations in case of the pristine system as well as bi-dopings in it with the 3$d$ late transition metal dopant atoms. The pristine systems are nonmagnetic and orbital hybridizations of the constituent atoms play significant role in stabilizing the respective morphologies. The doped systems are spin polarized. The magnetic couplings in the most stable structures are mostly anti-ferromagnetic in case of dopings in the sheet as well as cage morphologies which can be attributed to the {\it superexchange} interaction, while the optimal ferromagnetic and anti-ferromagnetic couplings are almost degenerate for the bi-dopings in the tube morphology. Separation between the two dopant atoms for the most preferred substitution, is found to be sensitive to the morphology of the host systems - the anti-ferromagnetic couplings between the two dopants in the sheet structure, favor short-ranged interaction, while the anti-ferromagnetic couplings are long ranged in case of the bi-dopings in the cage-shaped structure. Band gaps show wide variation for both the pristine as well as the doped systems and it shows overall decreasing trend while moving along sheet$\rightarrow$tube$\rightarrow$cage structures. The reduction in band gap variations with the decreasing aspect ratio, results mainly from the downward shifting of the conduction band minimum. Our results demonstrate that morphology variation of the semiconducting host system, will be very promising for band gap engineering in future optoelectronic applications.
\acknowledgments
We are grateful to Prof. T. Saha-Dasgupta for providing the computational facilities as well as for many simulating discussions. S. D. thanks Department of Science and Technology, India for support through INSPIRE Faculty Fellowship, Grant No. IFA12-PH-27.


\end{document}